\documentclass[twocolumn,showpacs,showkeys,amssymb,nobibnotes,superscriptaddress,aps,pra,bibnotes]{revtex4-1}

\usepackage{graphicx}% Include figure files
\usepackage{dcolumn}% Align table columns on decimal point
\usepackage{bm}% bold math
\usepackage[T1]{fontenc}
\usepackage{mathptmx}
\usepackage{mathrsfs}
\usepackage{physics}
\usepackage{simplewick}
\usepackage{mathrsfs}
\usepackage{amsthm,amsmath,amssymb}
\usepackage[bottom]{footmisc}
\usepackage{amssymb}
\usepackage{amsmath}
\usepackage{amsfonts}
\usepackage{bm}
\usepackage{graphicx}
\usepackage{subfigure}
\usepackage[dvipsnames]{xcolor}
\usepackage{calc}
\usepackage{epsfig}
\usepackage{color}
\usepackage[utf8]{inputenc}
\usepackage[colorlinks,citecolor=blue]{hyperref}

\newcommand{\be}{\begin{equation}}
\newcommand{\ee}{\end{equation}}
\newcommand{\bs}{\begin{split}}
\newcommand{\es}{\end{split}}

\begin{document}
\title{On the noise effect of test mass surface roughness in spaceborne gravitational wave detectors}

\author{Hao Yan}%
\affiliation{Center for gravitational experiment, MOE Key Laboratory of Fundamental Physical Quantities Measurements, The School of Physics, Huazhong University of Science and Technology, Wuhan 430074, China}

\author{Haixing Miao}%
\affiliation{State Key Laboratory of Low Dimensional Quantum Physics, Department of Physics, Tsinghua University, Beijing, China}

\author{Shun Wang}%
\affiliation{Center for gravitational experiment, MOE Key Laboratory of Fundamental Physical Quantities Measurements, The School of Physics, Huazhong University of Science and Technology, Wuhan 430074, China}

\author{Yiqiu Ma}%
\email{myqphy@hust.edu.cn}
\affiliation{Center for gravitational experiment, MOE Key Laboratory of Fundamental Physical Quantities Measurements, The School of Physics, Huazhong University of Science and Technology, Wuhan 430074, China}

\author{Zebing Zhou}%
\email{zzb@hust.edu.cn}
\affiliation{Center for gravitational experiment, MOE Key Laboratory of Fundamental Physical Quantities Measurements, The School of Physics, Huazhong University of Science and Technology, Wuhan 430074, China}
\date{\today}% It is always \today, today,

\begin{abstract}
Spaceborne gravitational wave detection mission has a demanding requirement for the precision of displacement sensing, which is conducted by the interaction between the laser field and test mass. However, due to the roughness of the reflecting surface of the test mass, the displacement measurement along the sensitive axis suffers a coupling error caused by the residue motion of other degrees of freedom. In this article, we model the coupling of the test mass residue random motion to the displacement sensing along the sensitive axis and derived an analytical formula of the required precision of the surface error for the spaceborne gravitational wave detectors. Our result shows that for the test masses in the LISA pathfinder, this coupling error will not contaminate the picometer displacement sensing. 
%Brownian thermal fluctuations 
\end{abstract}
%\keywords{test mass, surface error, laser interferometer, spaceborne gravitional wave detection}
\maketitle

\section{Introduction}
\begin{figure}[h]
\includegraphics[scale=0.25]{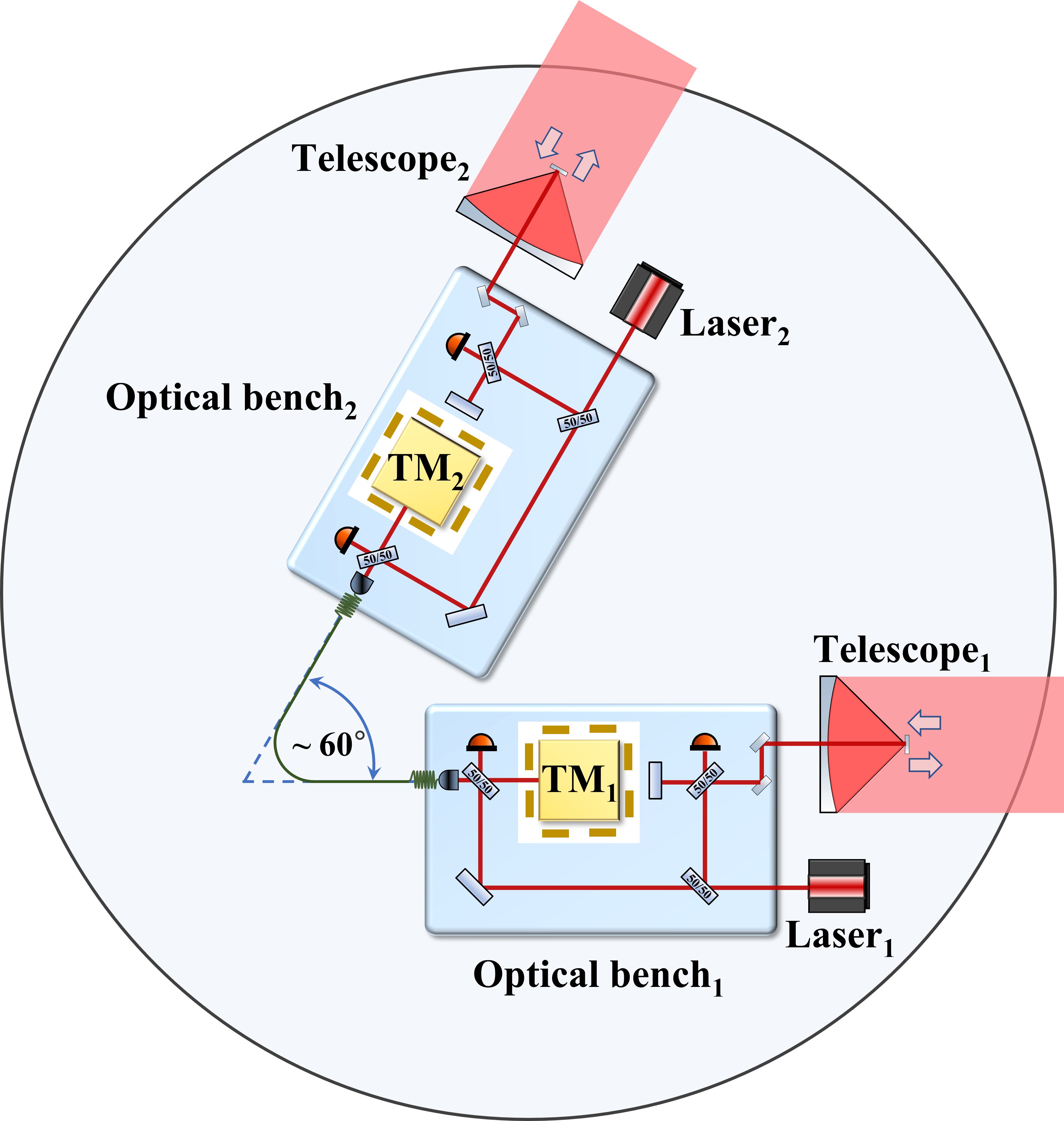}
\caption{\label{fig:payload} Schematic diagram of a satellite payload assembly in a spaceborne gravitational wave detector, where the laser field (ideally) couples to the test mass geodesic motion, which is preserved by the inertial sensor and drag-free controller.}
\end{figure}

Spaceborne laser interferometer gravitational wave detectors such as LISA\,\cite{Karsten1996}, Tianqin\,\cite{Tianqin}, and Taiji\,\cite{Taiji} are targetted on detecting the gravitational wave sources at the mili-Hertz range in the 2030s. The configuration of these gravitational wave detectors consists of three satellites that are connected by laser links. 
Each satellite consists of two independent subsystems in a 60-degree configuration, each of which contains an interferometry system (optical bench, laser source, telescope, etc.) and a gravitational reference sensor (GRS) with test mass inside, as shown in Fig.\,\ref{fig:payload}. Two key components inside these satellites are the test masses which serve as inertia references, and also the interferometer bench which measures the tiny displacement of the test mass. The detailed interaction process between the test mass and the laser field will affect gravitational wave detection.

%\R{Schematic Figure shows the inside components of a GW satellite.}

As an inertial reference, the cubic Au-Pt alloy test mass is surrounded by a conducting electrostatic shield with electrodes that are used for simultaneous capacitive position sensing and electrostatic force actuation\,\cite{GRS2003}. The drag-free and attitude control system (DFACS)\,\cite{lpf_PRL} 
ensures the very high dynamic stability of spacecraft and test masses, in particular along the sensitive axis along which the laser field gets reflected. For the other non-sensitive axis, the motion is controlled while introducing non-negligible disturbance noise, which causes beam walking and jitter effects by reflecting from the test mass surface as we shall discuss in this work. 

The surface roughness is an intrinsic property of the test mass originated from the manufacturing process\,\cite{Walsh99,Elson95,book2000,LYUKSHIN20211441,Butler2013}. On one hand, this surface roughness will introduce stray light and result in nonlinear errors in interferometry\,\cite{Marco2021,Sasso2019}, which have been well studied\,\cite{Church88,Vinet1996,Vinet1997,Ottaway:12,book1999}. On the other hand, the surface roughness will distort the wavefront of the reflected laser field. Furthermore, with the residue random motion perpendicular to the sensitive axis, such a wavefront distortion due to the surface roughness will not be time-independent, which could contribute noise to the phase measurement by the local interferometer in the satellite.  Analyzing such a noise effect is interesting since it couples several figures of merits of the satellite components such as the precision of the test mass manufacturing, interferometric phase measurement, and the residue motion of the inertia sensor.

This work is devoted to an in-depth analysis of how surface roughness contributes to displacement measurement in the spaceborne gravitational wave detector.  The interaction details of the residue test mass motion, surface roughness, and the light field are analyzed, and an analytical expression for the requirement of the surface roughness error given the picometer displacement sensing requirement is derived. Our results show that for the level of residue motion of the test masses in the LISA pathfinder, the surface roughness will not contribute significantly to the sensitivity of the picometer displacement sensing as long as the test mass manufacturing process limits the surface roughness error within about $100$\,nm. This is not a strict requirement with the current state-of-art technology.

\begin{figure}[h]
\includegraphics[scale=0.25]{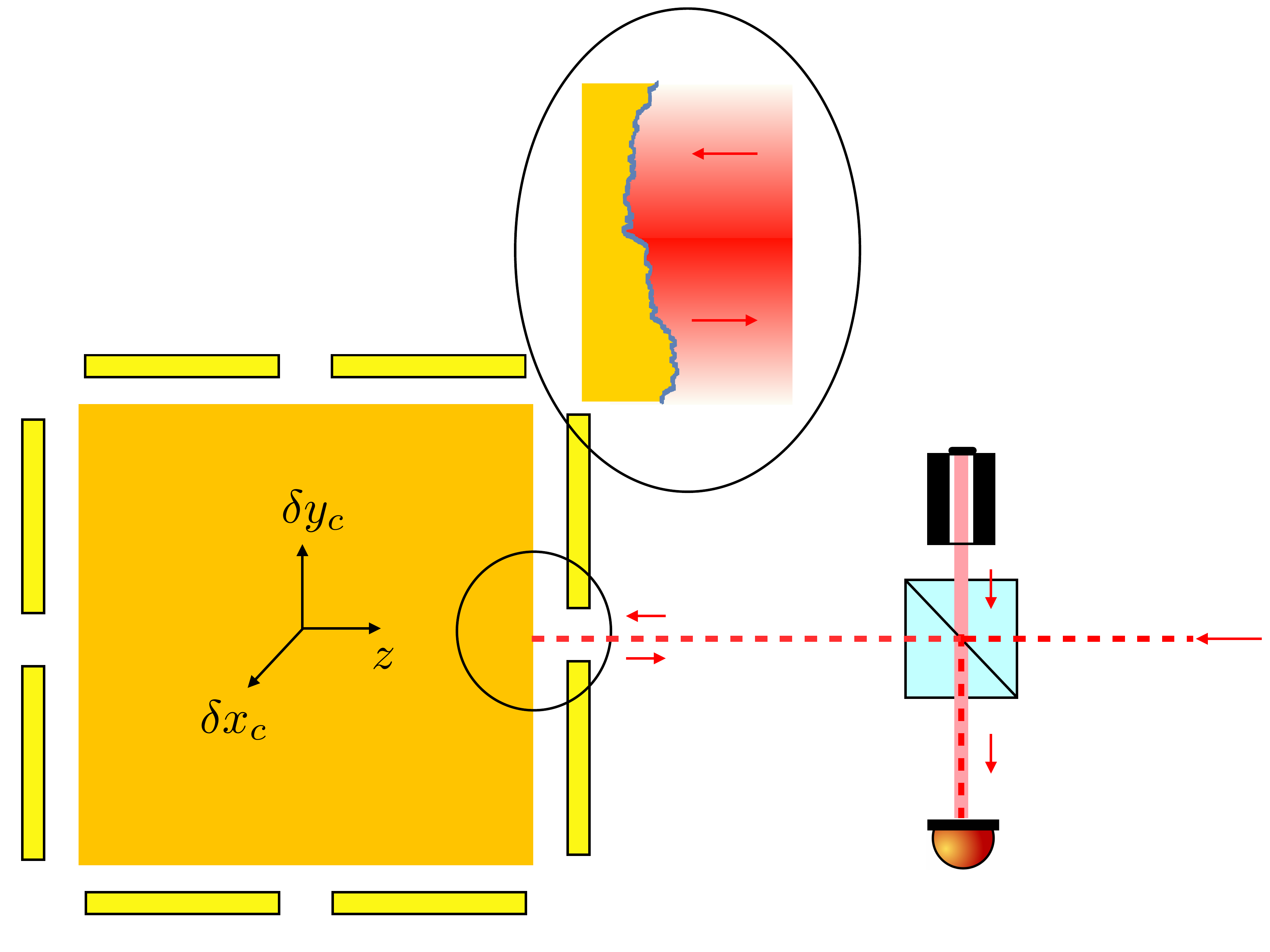}
\caption{\label{fig:1} Schematic diagram of the interferometric sensing of the test-mass displacement.}
\end{figure}

This paper is organized as follows. In Section II, theoretical modeling of the test mass surface roughness is presented. Section III discusses how it couples to the light field and distorts the wavefront. Then we discuss its impact on the interferometric phase measurement of the noise in Section IV. The summary and further discussions are presented in the last section.

\section{Modelling the surface roughness of a test mass}
The measured surface roughness is modelled by a two-dimensional height function $h(x,y)$, which is describes the random distribution of the surface height along the sensitive axis. In this work, we assume there is no cross-correlation between the $x$ and $y$ directions: $h(x,y)=h_x(x)h_y(y)$. The spatial correlation function the surface roughness is given by:
\be
C_{hh}(\mathbf{x}_\perp-\mathbf{x}'_\perp)=\overline{h(\mathbf{x}_\perp)h(\mathbf{x}'_\perp)},
\ee
where $\mathbf{x}_\perp=(x,y)$ and the overline represents the ensemble average. Therefore the correlation spectrum is:
\be
C_{hh}(\mathbf{k}_\perp)=\int d^2\mathbf{x}_\perp C_{hh}(\mathbf{x}_\perp)e^{i\mathbf{k}_\perp\cdot\mathbf{x}_\perp},
\ee
where $\mathbf{x}_\perp=(k_x,k_y)$. Since $h(x,y)=h_x(x)h_y(y)$, the correlation spectrum is also separable:
\be
C_{hh}(k_x,k_y)=\int dx e^{ik_x x}C^x_{hh}(x)\int dy e^{ik_y y}C^y_{hh}(y)
\ee
where $C^{x}_{hh}(x)=\overline{h_x(x)h_x(0)}$\,(the same for $C^{y}_{hh}(y)$). In our modelling, we assume that the surface roughness power spectral densities (PSD) of a test mass as:
\be\label{eq:PSD}
C^x_{hh}(k_x)=\frac{\mathcal{C}}{1+(k_x/k_\Lambda)^2},
\ee
where the $\mathcal{C}=\,{\rm 0.3\,\mu m^2}$ is the spectrum amplitude and the $k_\Lambda=1\,{\rm mm^{-1}}$ is the spectrum turning frequency. The effective surface roughness is\,\cite{khodnevych2020}
\be\label{eq:sigma1}
\begin{split}
\sigma_\lambda\approx&\left[\int^{1/\lambda}_{1/d}\int^{1/\lambda}_{1/d}dk_xdk_yC_{hh}(k_x,k_y)\right]^{1/2}\\
=&\xi\pi\mathcal{C}k_\Lambda,\\
\end{split}
\ee
where $\xi$ is a dimensionless coefficient that depends on the upper and lower limits of integration (in this paper, we take $\xi\approx1$). $\lambda$ is the measurement laser wavelength, and $d$ is the area of the test mass surface under measurement.

\begin{figure}[h]
\includegraphics[scale=0.4]{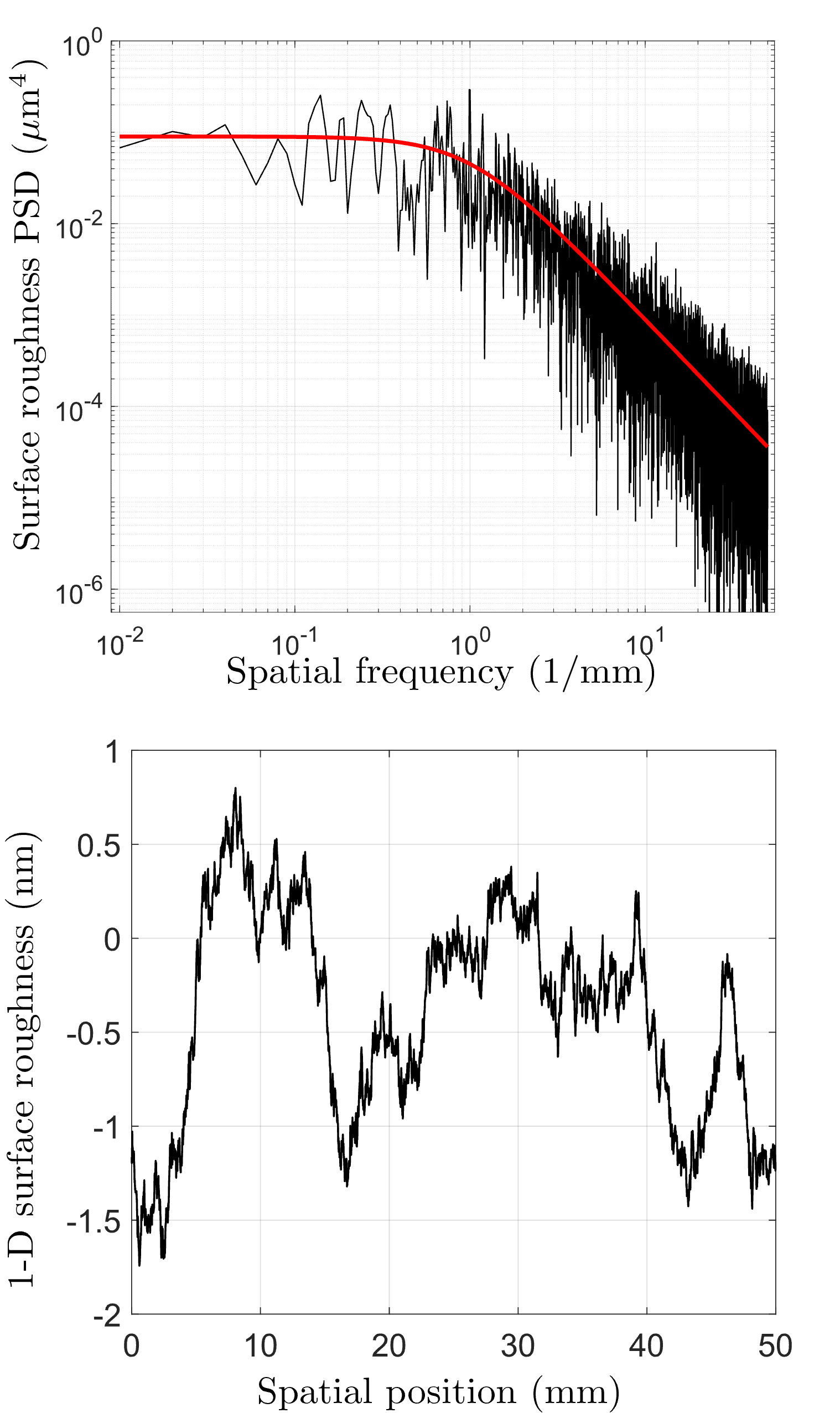}% Here is how to import EPS art
\caption{\label{fig:surface 1D} Upper panel: Power spectral densities of uncoated mirror surface roughness. The red line is the PSD model [see Eq.\,\eqref{eq:PSD}], and the black line is a simulated PSD created to generate the random maps used in our work. Lower panel: 1-D surface curve of a typical map. }
\end{figure}

\begin{figure}[h]
\includegraphics[scale=0.21]{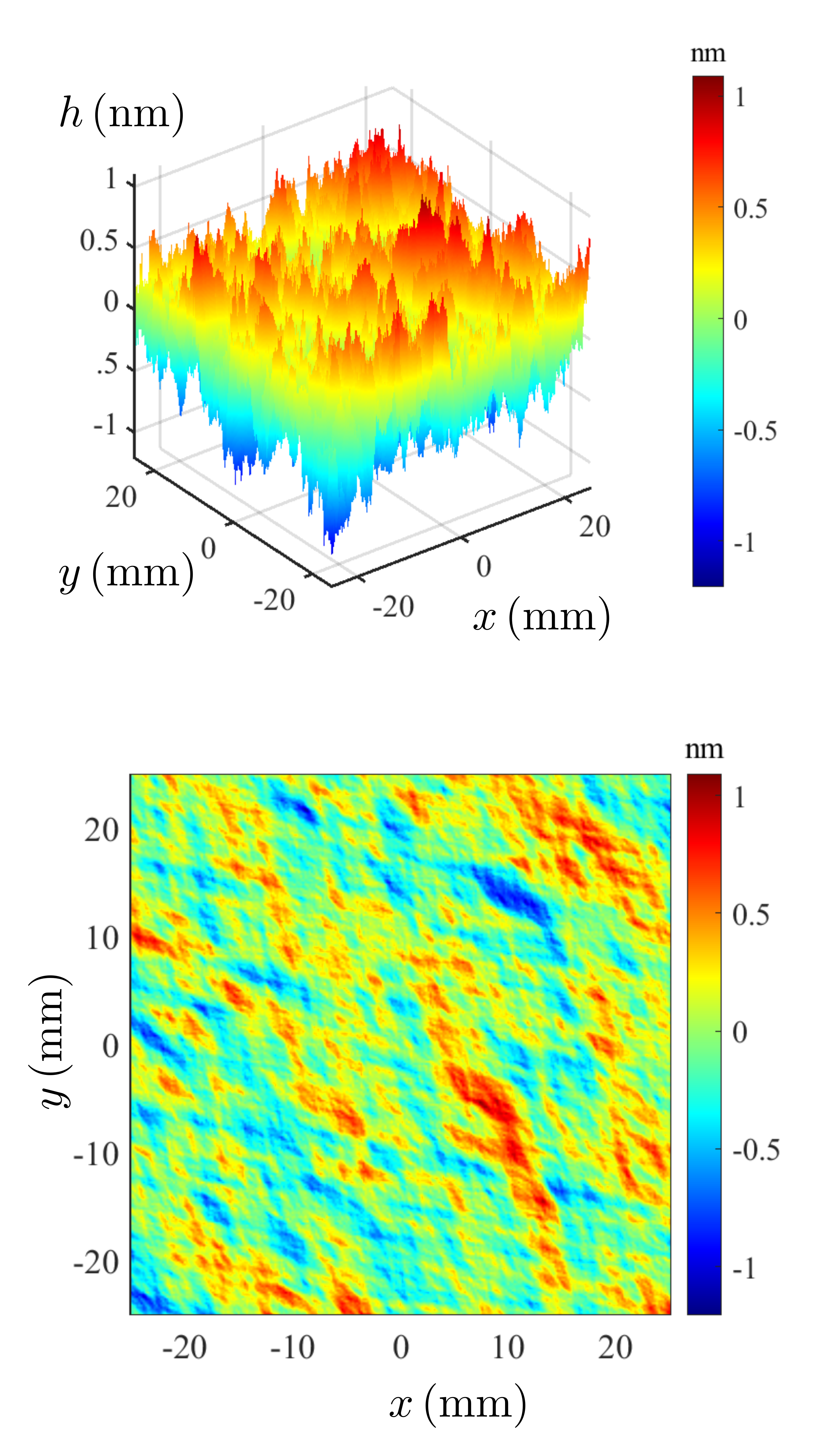}% Here is how to import EPS art
\caption{\label{fig:roughness 2D}Two-dimensional distribution of the surface roughness of the test masses.}
\end{figure}

We construct the surface roughness by multiplying the ASD (amplitude spectral density) by a random phase spectral density before transforming back to coordinate space to obtain the root-mean-square value, similar to\,\cite{Hong_T,Drori22}. Fig.\,\ref{fig:surface 1D} illustrates the surface roughness power spectral densities (PSD) and a typical surface curve. Considering the dimensions of the test mass (a few centimeters) and the diameter of the measurement laser beam (about one millimeter), we take the spatial wavelength in the range $ 25\,{\rm \mu m}-50\,{\rm mm}$ (see the horizontal coordinates of the power spectral density). Fig.\,\ref{fig:roughness 2D} illustrates a typical two-dimension surface map with a roughness rms of about $1\,{\rm nm}$ generated by the simulation.

%Consider the two-dimensional case similarly,
%\R{\be
%C^{xy}_{hh}(k_x,k_y)=\frac{\mathcal{C}^2}{\left[1+(k_x/k_\Lambda)^2\right]\left[1+(k_y/k_\Lambda)^2\right]},
%\ee}
%where the $\mathcal{C}_2=\,{\rm 1nm^2/mm^2}$ is the spectrum amplitude of two dimensions. 

In real manufacturing process, the surface roughness of the test mass manufactured by advanced machining technologies can be suppressed down to a few nanometers according to the current state-of-art.

\section{Light field scattered by the surface roughness}
\subsection{Hermite–Gaussian Modes}
\begin{figure}[h]
\includegraphics[scale=0.13]{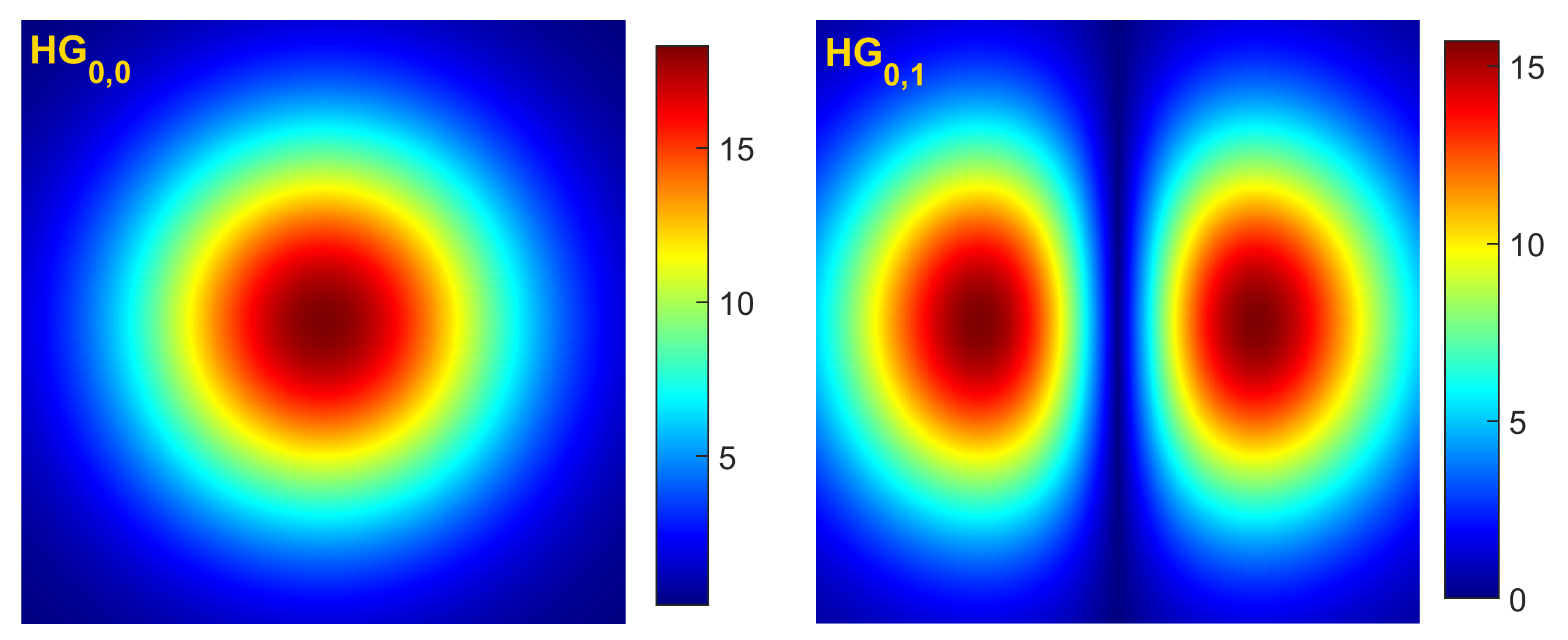} 
\caption{\label{fig:10} The Hermite Gaussian transversal spatial function of $\rm{HG}_{00}$ and $\rm{HG}_{01}$ modes.}
\end{figure}

The spatial electric field distribution of a freely-propagating light field is described by the Hermite-Gaussian mode:
\be\label{eq:HGmode}
\begin{split}
\frac{E_{nm}(x,y,z)}{E_0}=H_n\left(\frac{\sqrt{2}x}{w(z)}\right)H_m\left(\frac{\sqrt{2}y}{w(z)}\right){\rm exp}
\left[-\frac{x^2+y^2}{w^2(z)}\right]
\\
\frac{w_0}{w(z)}{\rm exp}
\left(-i\left[kz-(n+m+1){\rm arctan}\frac{z}{z_R}+\frac{k(x^2+y^2)}{2R(z)}\right]\right),
\end{split}
\ee
where $H_m(x)$ is the Hermitian function, $w(z)$ is the diameter of the cross-section of the light field as a function of propagation distance, $z_R$ is the Reighley length and $R(z)=\sqrt{w^2(z)+z_R^2}$. Two examples of the HG mode is plotted in Fig.\,\ref{fig:10}. All the Hermitian-Gaussian\,(HG) modes forms a Hilbert space and a general freely-propagating light field mode (normalized by $E_0$) can be expanded as:
\be
\begin{split}
&\psi(x,y,z)=\langle x,y,z|\psi\rangle=\sum_{j}a_j\langle x,y,z|{\rm HG}\rangle_j
\end{split},
\ee
where $|{\rm HG}\rangle_j$ is the eigenvector representing the HG mode with $j=(m,n)$ and $a_j$ is the expansion coefficient. The state-vector of the general light field is $|\psi\rangle=\sum_{j}a_j|{\rm HG}\rangle_j$. The $\langle x|{\rm HG}\rangle_j$ is the $j_{\rm th}$ HG mode in the coordinate representation, which is a compact form of Eq.\,\eqref{eq:HGmode}. The HG mode is a separable function and for notational simplicity later on, we will break the $x$-dependent part and $y$-dependent part of the HG mode by writing $|{\rm HG}_j\rangle=|{\rm HG}^x_j\rangle|{\rm HG}^y_j\rangle$

Usually the incoming signal field is assumed to be a pure HG mode with $n=m=0$. Reflecting by the imperfect test mass surface, the outgoing field will be a combination of all different HG modes where the expansion coefficient $a_j$ is determined by the mode shape. When the surface roughness effect is small, it only perturbs the outgoing field so that $|a_j|\ll|a_0|$ for $j\neq (0,0)$.
The phase measurement  by a heterodyne detector superpose the reflected light field with a local oscillator with $H_{00}(x)$ mode:
\be
\begin{split}
|\psi_m\rangle=&a_{\rm LO}|{\rm HG}\rangle_{00}+|\psi\rangle_{\rm ref}\\
=&a_{\rm LO}|{\rm HG}\rangle_{00}+a_{0}|{\rm HG}\rangle_{00}+\sum_{j\neq(0,0)}a_j|{\rm HG}\rangle_j
\end{split},
\ee
with the intensity which is proportional to the photo-electric current:
\be
I_m=\eta E_0^2(\psi_m|\psi_m)\approx\eta E_0^2|a_{\rm LO}+a_0|^2,
\ee
where the inner product here is defined as:
\be
(A|B)\equiv\int d^2x_\perp\langle A|x\rangle\langle x|B\rangle,
\ee
and the $\eta$ is the optical-electronical power conversion rate.
The integral is performed over the transversal $(x,y)$ plane. The physical meaning can be read out from dimensional analysis: the energy density flux is $\sim c\epsilon_0 |E|^2\propto (\psi_m|\psi_m)$, integration this energy density flux over the transversal plane leads to a quantity with dimension of power, which is the power of light absorbed by the photodetector.

\subsection{Dynamical coupling with the surface roughness}
The incoming field reflected from the surface dynamically couples to the test mass motion. The surface roughness $h(x,y)$ couples to the light field as:
\be
\psi_{\rm ref}(x,y,z)=r\psi_{\rm in}(x,y,z_0+h(x,y,t)),
\ee
where $r$ is the amplitude reflectivity and $h(x,y,t)$ is the surface roughness sensed by the laser field. This $h(x,y,t)$ is a random number since the transvesal motion is random. This coupling can be illustrated by the Gaussian mode under the paraxial approximation
\be
\begin{split}
\psi_{\rm ref}(x,y,z)=r[1-2ikh(x,y,t)]H_{00}(x,y,z_0)
\end{split}
\ee
where we have assumed the zero-th order flat surface located at $z=z_0$, and the small roughness allows the approximation ${\rm exp}[-2ikh(x,y)]\approx 1-2ikh(x,y)$. Moreover, the $h(x,y,z,t)H_{00}(x,y,z_0)$ can be expanded in terms of all HG normal modes.

During the heterodyne detection, as we have discussed in the Section III.A, 
there will be superposition of the local oscillator and the signal field and the final photo-electric current at the photo-detector is given by:
\be
I_m=\eta E_0^2(\psi_m|\psi_m)=\eta E_0^2\int d^2x_\perp \langle\psi_m|x_\perp\rangle\langle x_\perp|\psi_m\rangle,
\ee
where the trivial propagation in the z-direction is neglected for simplicity. The perturbation of the surface roughness to the measured photo-electric current $I_m$ is given as:
\be
\delta_h I_m=\eta2krE_0E_{\rm LO}\left(\frac{w_0}{w_D}\right)^2\int d^2 x_\perp {\rm exp}\left[-2\frac{x^2+y^2}{w^2_D}\right]h(x,y,t),
\ee
where $w_D$ is the diameter of the Gaussian optical beam at the detector position.

\begin{figure}[h]
\includegraphics[scale=0.35]{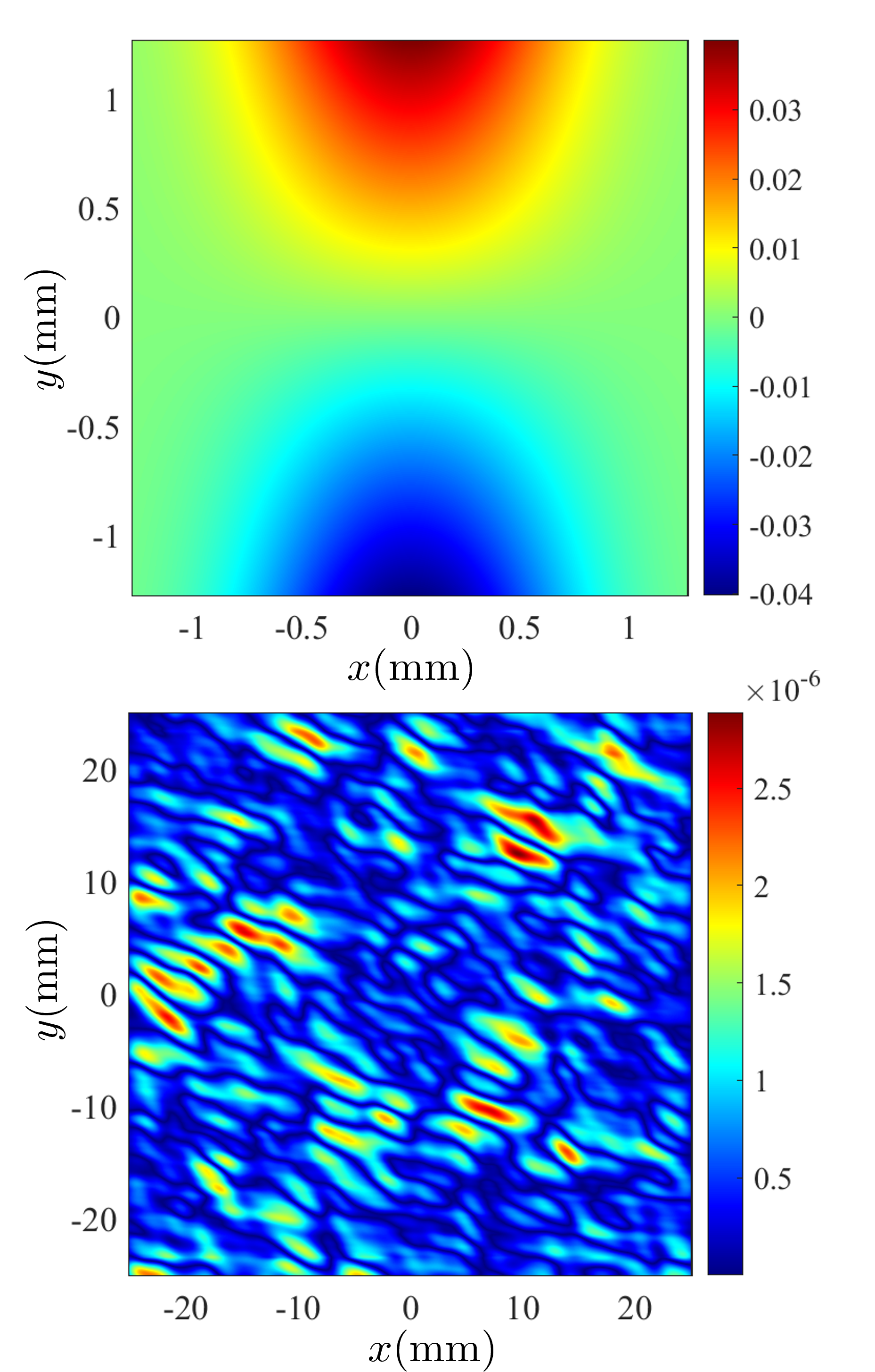}% Here is how to import EPS art
\caption{\label{fig:conversion_coefficient} Upper panel: filter function extended over the light spot. Lower panel: Two-dimensional distribution of the roughness-light conversion coefficient $|\langle {\rm HG}^y_{01}|h_y\rangle\langle {\rm HG}^x_{00}|h_x\rangle|$, where in the main text we only give the result for $\delta h_{\rm eff}$ of the geometric center of the test mass. The result is based on the original surface roughness distribution in Fig.\,\ref{fig:roughness 2D}.}
\end{figure}

The above result can be re-written using the effective optical path variation defined as:
\be\label{eq:heff(t)}
h_{\rm eff}(t)=\frac{4}{\pi w_D^2}\int d^2 x_\perp {\rm exp}\left[-2\frac{x^2+y^2}{w^2_D}\right]h(x,y,t),
\ee
so that $\delta_h I_m=\eta krE_0E_{\rm LO}h_{\rm eff}(t)\pi w_D^2/2$. For small transversal displacements, we have: 
\be
\begin{split}
&h(x,y,t)=h(x+\delta x_c(t))h(y+\delta y_c(t))\\
&\approx h_x(x)h_y(y)+h_x(x)\partial_yh_y(y)\delta y_c(t)+h_y(y)\partial_xh_x(x)\delta x_c(t),
\end{split}
\ee
where the first term $h_x(x)h_y(y)$ is a negligible time-independent DC source.
Therefore we have the coupling of the test mass transversal center of mass motion to the phase error as:
\be\label{eq:coupling_noise_1}
\begin{split}
\delta h_{\rm eff}(t)
=&\frac{4}{\pi w_D^2}\int d^2 x_\perp {\rm exp}\left[-2\frac{x^2+y^2}{w^2_D}\right]\\
&\left[h_x(x)\frac{\partial h_y(y)}{\partial y}\delta y_c(t)+h_y(y)\frac{\partial h_x(x)}{\partial x}\delta x_c(t)\right],
\end{split}
\ee
which has a simpler form after integration by parts:
\be
\begin{split}
\delta h_{\rm eff}(t)=&\langle {\rm HG}^y_{01}|h_y\rangle\langle {\rm HG}^x_{00}|h_x\rangle\delta y_{c}(t)\\
&+\langle {\rm HG}^x_{01}|h_x\rangle\langle {\rm HG}^y_{00}|h_y\rangle\delta x_{c}(t),
\end{split}
\ee
where the $\langle x|h\rangle\equiv h_x(x)$. Clearly, in this case the noise can be written as:
\be\label{eq:h_eff_spectrum}
\begin{split}
S_{h_{\rm eff}}(\Omega)=&|\langle {\rm HG}^y_{01}|h_y\rangle\langle {\rm HG}^x_{00}|h_x\rangle|^2S_{\delta y_{c}}(\Omega)\\
&+|\langle {\rm HG}^x_{01}|h_x\rangle\langle {\rm HG}^y_{00}|h_y\rangle|^2S_{\delta x_{c}}(\Omega).
\end{split}
\ee
Therefore the roughness-light conversion coefficient $\langle {\rm HG}^y_{01}|h_y\rangle\langle {\rm HG}^x_{00}|h_x\rangle$ is the key to obtain the displacement noise level along the sensitive axis. In Fig.\,\ref{fig:conversion_coefficient}, we present the absolute value of the conversion coefficient when the light spot is located at different positions of a surface with roughness distribution in Fig.\,\ref{fig:roughness 2D}. As we shall see later, this figure is very useful for estimate the test mass manufacture requirements. For a given test mass with the light spot precisely on the geometrical center of the test mass surface, what matters is the value at $(x=0,y=0)$, which is equal to $3\times 10^{-7}$ in the exemplary Fig.\,\ref{fig:conversion_coefficient}.  For LISA pathfinder \cite{LPF_platform,lpf_PRL}, the error of the transversal residue motion is $\delta x_c\sim 10^{-8}\,{\rm m}/\sqrt{{\rm Hz}}$, which corresponds to the displacement sensing noise around $\delta h_{\rm eff}\sim 3\times 10^{-15}\,{\rm m}/\sqrt{{\rm Hz}}$. Certainly it will not contribute a serious noise to the detector sensitivity.

\subsection{The surface roughness couples with imperfections of incoming light}
\begin{figure}[h]
\includegraphics[scale=0.1]{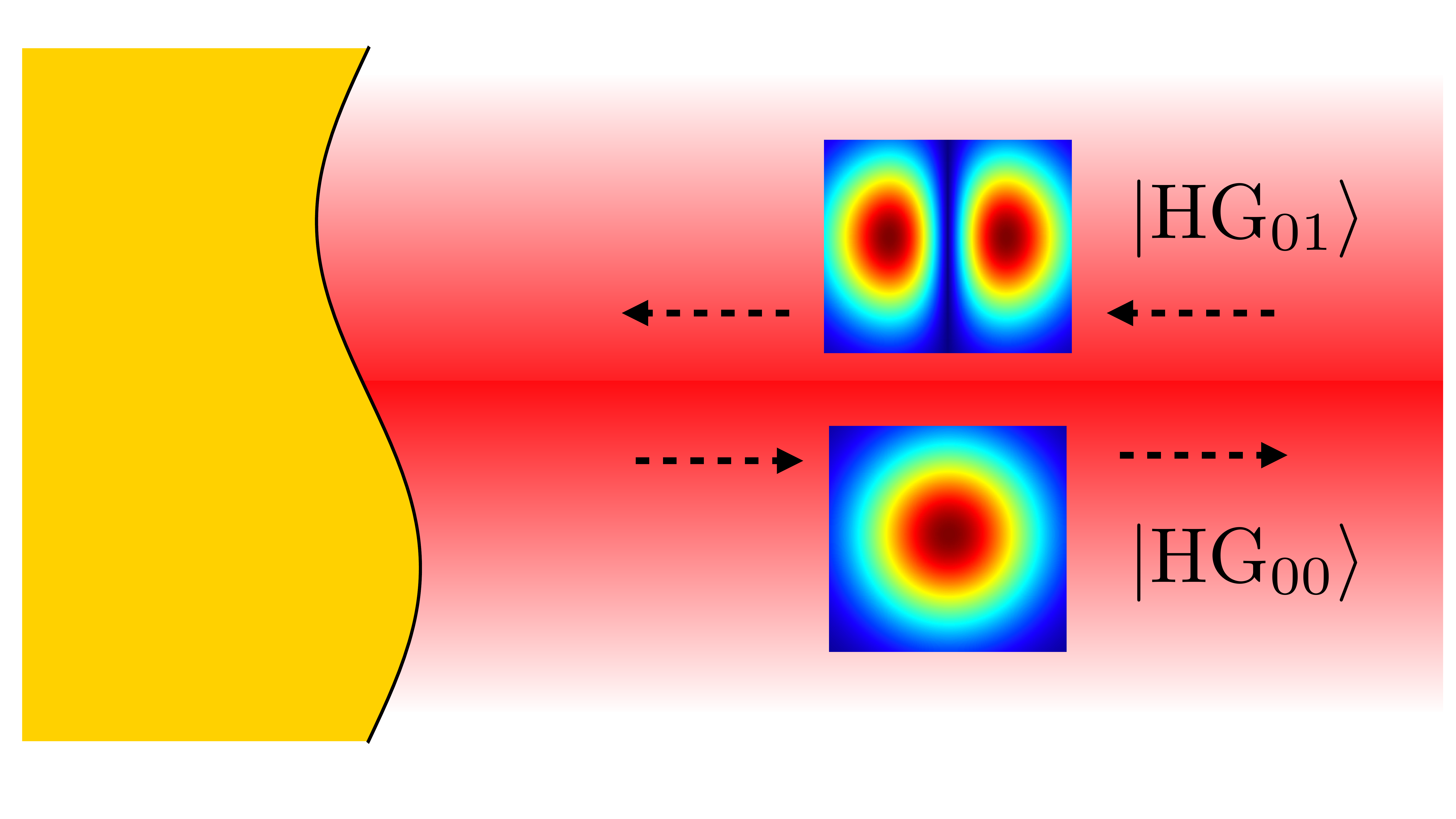}
\caption{Higher-order component in the incoming light field is converted back to the fundamental Gaussian mode via interaction with the surface roughness.}\label{fig:distortion_ref}
\end{figure}

Surface roughness has another effect on the phase error, which comes from the imperfections of the incoming light field. Let us suppose the incoming light contains other HG modes due to the distortion of the optical element in the optical path\,\cite{Mueller05,Barsotti_2010}:
\be
|\psi\rangle_{\rm in}=|{\rm HG}_{00}\rangle+\sum_{j\neq(0,0)}a_j|{\rm HG}_{j}\rangle.
\ee
Reflected from the rough surface, $|{\rm HG}_{01}\rangle$ can be converted back into the 
$|{\rm HG}_{00}\rangle$ mode in a stochastic way so that there will be a contamination of the phase measurement. Fig.\,\ref{fig:distortion_ref} shows schematically how the higher-order HG mode component in the incoming field being converted to $|{\rm HG}_{00}\rangle$ through reflecting by a rough surface.

The modulation of the surface roughness to the $|{\rm HG}_{01}\rangle$ in the incoming field can be written as:
\be
\sum_{j\neq(0,0)}a_j\langle x_\perp|{\rm HG}_{j}\rangle\rightarrow
\sum_{j\neq(0,0)}a_j(1+2ik\hat h)|{\rm HG}_{j}\rangle,
\ee
where the randomly perturbed z-directional roughness (denoted as operator $\hat h$) can have a non-zero overlap with the $|{\rm HG}_{00}\rangle$. This is important since the local oscillator of the heterodyne detector is a $|{\rm HG}_{00}\rangle$ mode, therefore its superposition with the perturbed field can be written as:
\be
a_{\rm LO}|{\rm HG}\rangle_{00}+ik\sum_{j\neq(0,0)}a_j\hat h|{\rm HG}\rangle_j,
\ee
and the corresponding perturbation to the photo-electric current is:
\be
\begin{split}
&\delta'_h I_m={\rm Re}\left[2ik\sum_{j\neq(0,0)}a_{\rm LO}^*a_j({\rm HG}_{00}|\hat h|{\rm HG}_j)\right]\\
&={\rm Re}\left[2ik\sum_{j\neq(0,0)}a_{\rm LO}^*a_j\int \frac{d^2x_\perp}{\pi w_D^2}\langle {\rm HG}_{00}|x_\perp\rangle h(x,y)\langle x_\perp|{\rm HG}_j\rangle\right].
\end{split}
\ee

We can similarly construct an effective optical path variation in this case, defined as:
\be
\begin{split}
&\delta'h_{\rm eff}=\\
&{\rm Re}\left[2ik\sum_{j\neq(0,0)}a_{\rm LO}^*a_j\int\frac{d^2x_\perp}{\pi w_D^2}\langle {\rm HG}_{00}|x_\perp\rangle h(x,y)\langle x_\perp|{\rm HG}_j\rangle\right].
\end{split}
\ee
The time-dependence of $\delta'h_{\rm eff}$ may have two origins. (1) The imperfections of the optical element in the interferometer may be time-dependent so that $a_j=a_j(t)$, which directly couples with a time-independent $h(x,y)$ roughness, contributing to the phase error:
\be
\begin{split}
&\delta^{a_1}h_{\rm eff}(\Omega)=\\
&{\rm Re}\left[2ik\sum_{j\neq(0,0)}a_{\rm LO}^*a_j(\Omega)\int\frac{d^2x_\perp}{\pi w_D^2}\langle {\rm HG}_{00}|x_\perp\rangle h(x,y)\langle x_\perp|{\rm HG}_j\rangle\right].
\end{split}
\ee
Supposing the HG$_{01}$ mode dominates the random optical imperfections,  the noise spectrum can be written as:
\be\label{eq:S_a1a1}
S^{a_1}_{h_{\rm eff}}(\Omega)
=\left|\int \frac{d^2x_\perp}{\pi w_D^2}\langle {\rm HG}_{00}|x_\perp\rangle h(x,y)\langle x_\perp|{\rm HG}_{01}\rangle\right|^2S_{a_1a_1}(\Omega).
\ee
With this formula, the picometer displacement sensing accuracy will set a requirement to the imperfections of the incoming light field, as we will discuss in the next section.

\begin{figure}[h]
\includegraphics[scale=0.35]{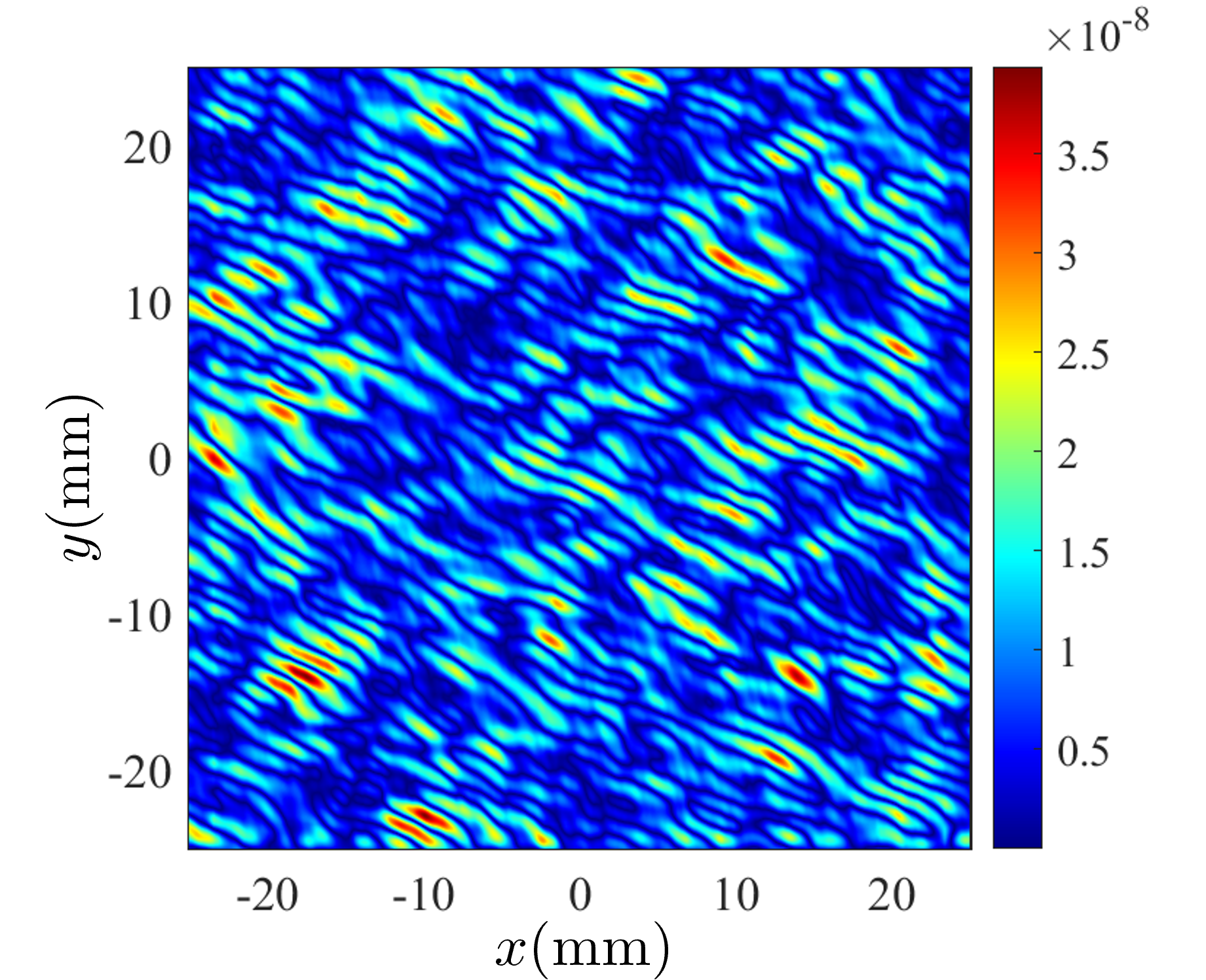}
\caption{\label{fig:coefficient_a_1}  Conversion coefficient between the surface roughness and the imperfections of incoming light $\sqrt{S_{h'_{\rm eff}}/a^2_1S_{\delta x_c}}$.}
\end{figure}

(2) Similar to the subsection B, the $h(x,y)$ is fluctuation due to the transversal residue center of mass motion of the test mass:
\be
\begin{split}
\delta'h_{\rm eff}(\Omega)={\rm Re}&\left[2ik\sum_{j\neq(0,0)}a_{\rm LO}^*a_j\int \frac{d^2x_\perp}{\pi w_D^2} \langle {\rm HG}_{00}|x_\perp\rangle\langle x_\perp|{\rm HG}_j\rangle\right.\\
&\left.\frac{\partial h(x,y)}{\partial y}\delta y_c(\Omega)+\frac{\partial h(x,y)}{\partial x}\delta x_c(\Omega)\right],
\end{split}
\ee
with the corresponding noise spectrum as:
\be
\begin{split}
S_{h'_{\rm eff}}&(\Omega)
=\\
&\frac{2a^2_1}{w^2_D}\left|\int \frac{d^2x_\perp}{\pi w_D^2}
\langle {\rm HG}_{00}|x_\perp\rangle\langle x_\perp|{\rm HG}_{01}\rangle\frac{\partial h(x,y)}{\partial y}\right|^2S_{\delta x_c}(\Omega)\\
+&\frac{2a^2_1}{w^2_D}\left|\int \frac{d^2x_\perp}{\pi w_D^2}
\langle {\rm HG}_{00}|x_\perp\rangle\langle x_\perp|{\rm HG}_{01}\rangle\frac{\partial h(x,y)}{\partial x}\right|^2S_{\delta y_c}(\Omega),
\end{split}
\ee
where the conversion coefficient is plotted in Fig.\,\ref{fig:coefficient_a_1}. Since the interferometer in the GW satellite has a precise control of the beam, the $a_j$ is typically small thereby the $\delta'h_{\rm eff}$ should be a secondary effect compared with the first one. 

\section{Noise level and Test mass manufacture requirements}
After analyzing the details of the coupling between the light field and the surface roughness, this section devotes to discussing the noise level and the corresponding requirements of the test mass manufacture. This discussion is divided into two parts: (1) for a given test mass with a fixed and known surface roughness distribution, we want to estimate the displacement sensing noise level; (2) for a general manufacturing process, which corresponds to an ensemble of surface roughness realizations, we want to discuss how the noise level estimation put a requirement on the test mass manufacturing process.

\begin{figure}[h]
\includegraphics[scale=0.18]{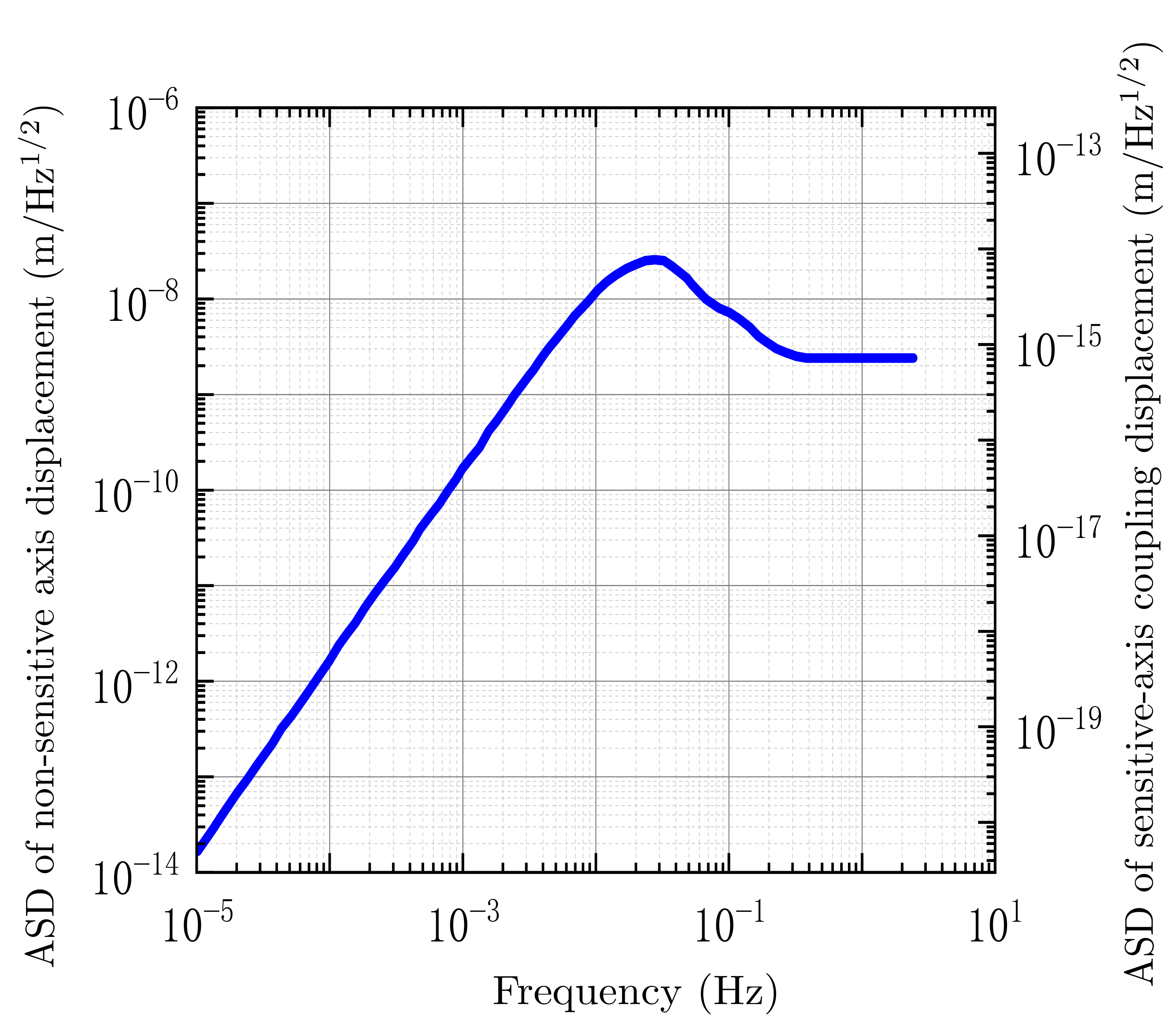}% Here is how to import EPS art
\caption{\label{fig:LPF_yz} Amplitude spectral densities (ASD) of non-sensitive axis displacement of LISA Pathfinder \cite{LPF_platform}.}
\end{figure}

Firstly, for a given test mass with known surface roughness distribution $h(x,y)$, it is straightforward to calculate the phase error through the spectrum of $\delta h_{\rm eff}$, $\delta h^{a_1}_{\rm eff}$, and $\delta h'_{\rm eff}$, which is shown in Fig.\,\ref{fig:LPF_yz}.

Secondly, for setting a requirement on the test mass manufacturing process, we should assume an ensemble of test mass realization to mimic the uncertainty of the manufacturing process. Each test mass in this ensemble interacts with the laser field where the light spot is on the geometrical center of one surface of the test mass cubic. Since these ensemble test masses have a different realization of the surface roughness, the reflected light field will be modulated in a different and stochastic way. To give an indicator of surface manufacture, we need to compute an ensemble average of the phase error over all surface roughness realizations. According to the \emph{ergodic theorem}, this ensemble average is equivalent to the average over different spot center points on a sufficiently large surface plane, which allows us to make use of the result given in the previous section.

The error calculated using ergodic theorem mathematically can be represented as:
\be\label{eq:heff_variance}
\begin{split}
\overline{\delta h^2_{\rm eff}(\Omega)}&=\frac{1}{N}\sum_j\delta h^2_{\rm eff}[\Omega, h_j(x_{\rm spot})]\\
&=\frac{1}{A}\int d^2x_{\perp c}\delta h^2_{\rm eff}[\Omega, h_j(x_{\perp c})],
\end{split}
\ee
where $x_{\perp c}$ is the coordinate of the light spot center at the surface plane, and $\delta h_{\rm eff}[\Omega, h_j(x_{\perp c})]$ is already obtained in Fig.\,\ref{fig:conversion_coefficient}. The gravitational wave detection mission requires that:
\be
\Delta \delta h_{\rm eff}(\Omega)=\sqrt{\overline{\delta h^2_{\rm eff}(\Omega)}}\leq 1\,{\rm pm}/\sqrt{{\rm Hz}},
\ee
from which a requirement on the surface roughness can be derived as follows.

Let us take the noise in Eq.\,\eqref{eq:coupling_noise_1} as an example. Substituting Eq.\,\eqref{eq:coupling_noise_1} into Eq.\,\eqref{eq:heff_variance} and after some straightfoward algebra, the result can be written in terms of the relationship among the displacement sensing error, surface roughtness spectral amplitude and the test mass residue motion error:
\be
\overline{\delta h^2_{\rm eff}}(\Omega)=\mathcal{C}^2H_{\delta x_c}\overline{\delta x^2_c}(\Omega)+\mathcal{C}^2H_{\delta y_c}\overline{\delta y^2_c}(\Omega),
\ee
where the coefficients $H_{\delta x_c}$ and $H_{\delta y_c}$ are given by:
\be
\begin{split}
H_{\delta x_c}=&\int\frac{d^2x_{\perp c}}{A}d^2k_\perp\frac{2}{\pi w_D^2}\frac{1}{1+(k_x/k_\Lambda)^2}\frac{1}{1+(k_y/k_\Lambda)^2}
\\
&|\mathcal{F}_{k_x}[\langle x|{\rm HG}^x_{01}(x_c)\rangle]\mathcal{F}_{k_y}[\langle y|{\rm HG}^y_{00}(y_c)\rangle]|^2\\
H_{\delta y_c}=&\int\frac{d^2x_{\perp c}}{A}d^2k_\perp\frac{2}{\pi w_D^2}\frac{1}{1+(k_x/k_\Lambda)^2}\frac{1}{1+(k_y/k_\Lambda)^2}
\\
&|\mathcal{F}_{k_y}[\langle y|{\rm HG}^y_{01}(y_c)\rangle]\mathcal{F}_{k_x}[\langle x|{\rm HG}^x_{00}(x_c)\rangle]|^2,
\end{split}
\ee
in which $\mathcal{F}_{k}[...]$ is the Fourier transformation defined as:
\be
\mathcal{F}_{k}[f(x)]\equiv\int dx e^{ik x}f(x),
\ee
and $|{\rm HG}^x_{00/01}(x_c)\rangle$ means the HG modes $|{\rm HG}^x_{00/01}\rangle$ centered at $x_c$. 

\be
\begin{split}
H_{\delta x_c}=H_{\delta y_c}=&\frac{\sqrt{2\pi}k_\Lambda^3}{w_D}e^{-\frac{k_\Lambda^2w_D^2}{2}}{\rm Erfc}(\frac{k_\Lambda w_D}{\sqrt{2}})\\
&\left[1-\sqrt{\frac{\pi}{2}}k_\Lambda w_De^{-\frac{k_\Lambda^2w_D^2}{2}}{\rm Erfc}(\frac{k_\Lambda w_D}{\sqrt{2}})\right],
\end{split}
\ee
where ${\rm Erfc}(...)$ is the complementary error function. Finally we have:
\be
\sigma_\lambda=\xi\pi\mathcal{C}k_\Lambda\leq\left[\frac{k_\Lambda^2\overline{\delta h^2_{\rm eff}}(\Omega)}{H_{\delta x_c}[\overline{\delta x^2_c}(\Omega)+\overline{\delta y^2_c}(\Omega)]}\right]^{1/2},
\ee
or equivalently
\be
\overline{\delta x^2_c}(\Omega)+\overline{\delta y^2_c}(\Omega)\leq\frac{k_\Lambda^2\overline{\delta h^2_{\rm eff}}}{\pi^2\sigma^2_\lambda H_{\delta x_c}}.
\ee
These formulae set a constraint on the surface roughness once the displacement sensing error and the transversal residue motion error are given, or a constraint on the precision for the residue motion control on the insensitive axis with a given surface roughness.  For the space-borne gravitational wave detection mission, the LISA pathfinder residue motion $\delta x_c\sim\delta y_c\sim 10^{-8}\,{\rm m}/\sqrt{{\rm Hz}}$\,\cite{LPF_platform}, the surface roughness error must be lower than $100\,{\rm nm}$ with $H_{\delta x_c}\sim 10^{12}\,{\rm m^{-4}}$.  Moreover, for the fluctuation due to the imperfections of the incoming light field, the formula Eq.\,\eqref{eq:S_a1a1} leads to a requirement for the jitter of the incoming light, where we have $\sqrt{S_{a_1a_1}}<10^{-2}/\sqrt{{\rm Hz}}$ for a picometer displacement sensing accuracy.

\section{Discussion and Conclusions}
In this work, we have investigated the effects of test mass surface roughness in the spaceborne gravitational wave detectors, which was not shown much in the previous works. It is worth mentioning that for ground-based gravitational wave detectors such as LIGO/VIRGO, the surface roughness of the suspended mirrors also impacts the detector sensitivity, which has been studied in many works\,\cite{Bondarescu2008,Bondarescu2006,Walsh992,Ast2021,Tao21}. A comparison between the surface roughness effect on space-borne and ground-based detectors is worth a discussion. 

In the ground-based gravitational wave detector, two test mass mirrors form a Fabri-Perot arm cavity, which enhances the detector's sensitivity. The detector sensitivity is dominated by quantum noise at the interested frequency band, and it is quantum optically enhanced by injecting the squeezed light\,\cite{Aasi2013,SCHNABEL20171}. Therefore achieving a full-coherent cavity is the key to the quantum metrology in the ground-based detector. Surface roughness can scatter the cavity field from the fundamental Gaussian mode into other higher-order modes, thereby contributing to an optical loss
that can degrade the coherence of the quantum light field. These effect has been discussed by\,\cite{Drori22}. Moreover, the mirror surface roughness can also affect the interferometer designs based on injecting high-order mode laser, which targets mitigating the thermal noise effect. In these designs, it is found that the optical scattering by the mirror surface roughness will seriously affect the interferometer operating with higher-order Laguerre-Gaussian laser modes, while higher-order Hermit-Gaussian fields could be useful with future designs\,\cite{Ast2021,Heinze2022}. In summary, ground-based detectors are more concerned about the effect of mirror surface roughness on static optical loss.

Similarly, note that the wavefront distortion generated by the surface scattering also contributes to a scattering loss as in the ground-based detector case. However, since the interferometric sensing of the test mass motion in a LISA-type space-borne detector does not involve the optical cavity structure and quantum metrology, such a scattering loss will not alter the shot noise floor. Differently, our work emphasized the dynamic side of this effect. The residue random motion of the test mass along the non-sensitive axis couples to the surface roughness, and finally affects the laser field along the sensitive direction or the fluctuating component of the incoming field that couples to the Gaussian mode via surface roughness. 

In this paper, we have discussed the noise mechanism due to the dynamic coupling of the test mass surface roughness with the light field. Our conclusion is that with the residue motion of the test mass along the non-sensitive axis in the LISA pathfinder, the nano-meter scale surface roughness only contributes a $\sim 0.01\,{\rm pm}/\sqrt{\rm Hz}$ noise level to the displacement sensing along the sensitive axis at around $10\,{\rm mHz}$. As long as the test mass manufacturing state of the art can reduce the surface roughness below the 100\,nm level, a $10\,{\rm nm}/\sqrt{\rm Hz}$ residue motion would not significantly affect the gravitational wave detection.

\acknowledgements
We thank Dr. Ting Hong for useful discussions. H. Y.\,is supported by National Natural Science Foundation of China (Grant No. 12105375). Y. M.\,is supported by the start-up funding provided by Huazhong University of Science and Technology. S. W. is supported by National Natural Science Foundations of China (Grant No. U20A2077) and National Key R\&D Program of China (2021YFC2202300).  H. M.\,is supported by the State Key Laboratory of Low Dimensional Quantum Physics and the start-up fund from Tsinghua University. Z. Z.\,is supported by National Natural Science Foundations of China (Grant Nos. 11727814 and 11975105).

\nocite{*}
\bibliography{reference}% Produces the bibliography via BibTeX.
\end{document}